\documentclass[pdflatex,sn-mathphys-num, iicol]{sn-jnl}
\usepackage{graphicx}%
\usepackage{multirow}%
\usepackage{amsmath,amssymb,amsfonts}%
\usepackage{amsthm}%
\usepackage{mathrsfs}%
\usepackage[title]{appendix}%
\usepackage{xcolor}%
\usepackage{textcomp}%
\usepackage{manyfoot}%
\usepackage{pdfpages}
\usepackage{slashed}
\usepackage{booktabs}%
\usepackage{algorithm}%
\usepackage{algorithmicx}%
\usepackage{algpseudocode}%
\usepackage{listings}%
\usepackage{tabularx}
\theoremstyle{thmstyleone}%

\usepackage{tabularx,array}
\theoremstyle{thmstyletwo}%
\usepackage{longtable}
\newcolumntype{Y}{>{\raggedright\arraybackslash}X}
\theoremstyle{thmstylethree}%
\usepackage{indentfirst}
\begin{document}
\title[Scalaron-driven Dark Matter during Warm Inflation via UV Freeze--in]{Scalaron-driven Dark Matter during Warm Inflation via UV Freeze--in}
\author*[1]{\fnm{Florian} \sur{Millo}}\email{florian.millo@cnrs.fr}
\affil*[1]{\orgdiv{Physics Department}, \orgname{Sorbonne Universit\'e (CNRS)}, \orgaddress{\street{4 place Jussieu}, \city{Paris}, \postcode{75005}, \country{France}}}


\abstract{We investigate the production of Dark Matter (DM) within the warm Higgs--Starobinsky (HS) inflation. By adopting the ultraviolet (UV) freeze--in mechanism -- where the DM number density is initially negligible but is populated over time through non-renormalizable interactions of defined mass dimensions -- we find that the DM production within the HS potential has a characteristic timing. For both DM masses, $m_\chi=1$ MeV and $m_\chi=100$ GeV, the yield of DM occurs close to the termination of inflation and its transition to the radiation dominated (RD) era. The present--day DM relic abundance is generated for both DM masses with non-renormalizable operators of mass dimension $D=\{8,9\}$ within strong dissipation regimes. The obtained results suggest that the form of the HS potential and the associated scalaron dynamics favor UV freeze--in DM production as the scalaron approaches the minimum of the potential, where energy transfer to the thermal bath becomes particularly efficient. This distinctive transition behavior differentiates the HS model from other inflationary models and advances our understanding of DM production.\\ \\ Published in Eur. Phys. J. C 86, 910 (2026). DOI: 10.1140/epjc/s10052-026-16169-y}


\maketitle

\section{Introduction}
Dark Matter (DM) is a form of non-baryonic matter that is hypothesized to have been present since the primordial moment of the Universe. While we know from astrophysical observations that DM comprises up to 95\% of the matter content of the Universe \cite{planck_collaboration_planck_2020, jarosik_seven-year_2011}, little is known about how its production mechanism works or how its present-day abundance is generated. A promising avenue for explaining DM abundance involves interactions between DM particles (in the hidden or visible sectors \cite{cheung_origins_2011}) and particles in a thermal bath. Two efficient mechanisms called freeze--out and freeze--in have been extensively employed to understand the DM abundance \cite{bernal_dawn_2017,elahi_ultraviolet_2015, hall_freeze-production_2010, cheung_origins_2011}. In freeze--out, the DM is in thermal equilibrium with the bath at early times and its abundance evolves within the equilibrium distribution until it decouples from the bath \cite{elahi_ultraviolet_2015}. On the other hand, in freeze--in the DM never achieves a thermal equilibrium with the bath \cite{hall_freeze-production_2010, elahi_ultraviolet_2015}. In the freeze--in mechanism, the DM number density is initially negligible but can evolve over time through: renormalizable interactions within a temperature independent mechanism known as infrared (IR) freeze--in setting; or through non-renormalizable interactions within a temperature dependent mechanism known as ultraviolet (UV) freeze--in setting \cite{elahi_ultraviolet_2015, hall_freeze-production_2010}. In addition, for the DM number density to be initially negligible, the hidden and visible sectors must be decoupled (or feebly) coupled at all the times \cite{elahi_ultraviolet_2015}. Moreover, the UV freeze--in setting is highly sensitive to the early times  -- especially to the high temperatures and how radiation is produced -- making its examination within a warm inflationary period both natural and well-motivated, since these conditions are precisely realized in a warm inflationary era. The Warm Inflation (WI) framework \cite{berera_warm_1995} provides a mechanism for a gradual shift from the warm inflationary era to the radiation era via a dissipation that depends on temperature. This framework can be intimately connected to the UV freeze--in mechanism for DM production.\\
\indent Recently, in Ref.~\cite{freese_dark_2024}, the authors present a theoretical framework merging UV freeze--in with warm inflation for the production of DM relic abundance. They adopted a quartic potential $(V(\varphi)\sim\varphi^4)$ which, while incompatible with cosmic microwave background (CMB) constraints in cold inflation, becomes viable through the dissipative mechanism of warm inflation. Moreover, in Ref.~\cite{wang_purely_2025}, the authors use the same mechanism to explore the gravitational DM production within a quadratic potential energy $(V(\varphi)\sim\varphi^2)$. These potentials are well suitable for the direct decay mechanism of 2--to--2 scattering of the Standard Model (SM) particles. 
A particularly interesting possibility is the Higgs--Starobinsky (HS) potential, in which a scalaron particle drives the cosmic inflation \cite{samart_warm_2022, bernal_uv_2020}. An appealing scenario is to implement the UV freeze--in mechanism for DM production by scalarons. As far as we are aware, this scenario rests unexplored. \\
\indent In this paper, we investigate the scalaron-driven DM production during warm inflation via UV freeze--in mechanism. The key assumption of this work is that the 2--to--2 scattering processes between Standard Model (SM) particles are permitted to occur in the HS potential. Once we have settled this assumption, our investigation is similar to the one performed in Ref.~\cite{freese_dark_2024}. Our main result is that the DM production occurs close to the transition between inflation and radiation dominated era. The present–day DM relic abundance is generated for both DM masses $m_\chi=\{1\,\mathrm{MeV}, 100\,\mathrm{GeV}\}$ through non-renormalizable interactions of mass dimension $D = \{8, 9\}$. In other words, the scalaron decays through the UV freeze--in mechanism and produces enough DM relic abundance at inflation-to-radiation transition through operators of mass dimension $D=\{8,9\}$. We want to emphasize that a tentative study of scalaron-driven DM has been performed in Ref.~\cite{bernal_uv_2020} but the authors did not calculate the yield of the DM relic abundance. They follow a rather methodological form of showing the asymptotics of the HS model in which a DM mass boundary larger than a few keV was deduced, and the non-renormalizable interactions must be of mass dimension $D<10$. Our pivotal contribution lies in calculating the DM yield within a warm HS potential and discussing the consequences of the cutoff scale. An attractive coincidence of scale is found that relates the calculated cutoff scale with the origin of the neutrino mass.\\
\indent This manuscript is organized as follows. In Sec.~\ref{secWIdyn}, we write the general WI dynamic equations and show the evolution of the dynamic quantities. Next, in Sec.~\ref{secDMyield}, we implement the Boltzmann equation to deduce the DM yield for different DM masses. In Sec.~\ref{natureDM}, we discuss the nature of the DM particle. In Sec.~\ref{discSumm}, we discuss and summarize our findings. \\
\indent Throughout this manuscript, the following conventions are adopted: Friedmann--Lema\^itre--Robertson--Walker (FLRW) metric $g_{\mu\nu}dx^\mu dx^\nu=-dt^2+a^2(t)d\vec{x}^2$ is used for a flat and isotropic Universe where $a(t)$ is the scaling factor and $c=1$. The reduced Planck mass is $ M_{\mathrm{Pl}} \approx 1/\sqrt{8\pi G}\approx2.4\times10^{18}$ GeV. 



\section{Warm Inflation Dynamics}\label{secWIdyn}
\begin{figure}[!t]
    \centering
    \includegraphics[width=1\linewidth]{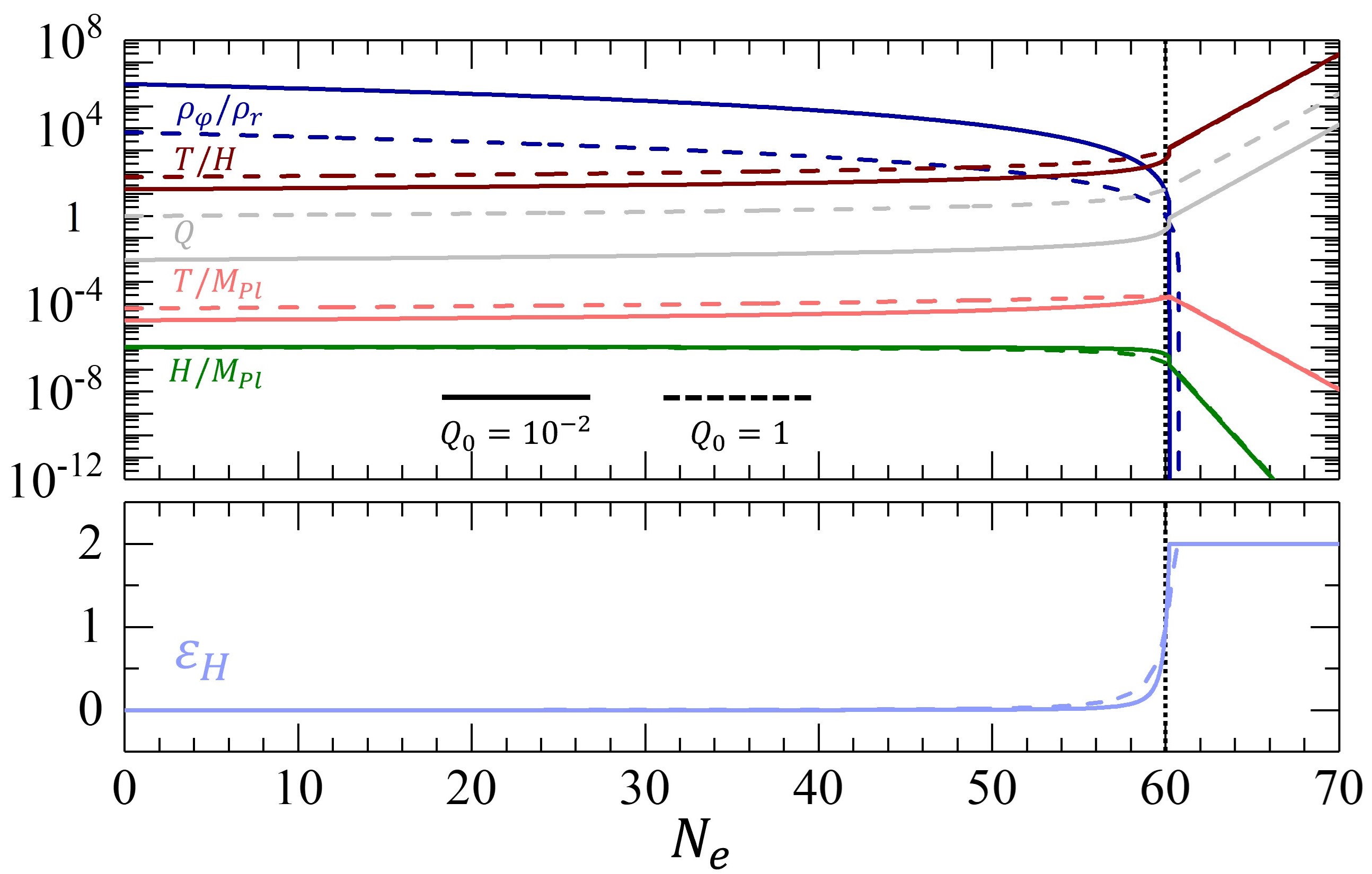}
    \caption{Evolution of dynamic quantities within the WI under HS potential [Eq.~\ref{HSPotential}] with scalaron mass $M\approx 2.13 \times10^{-6}  M_{\mathrm{Pl}}$. For the dissipation mechanism, two initial values are assumed for strong regime $Q_0=1$ (dashed) and weak regime $Q_0=10^{-2}$ (solid). The vertical dashed line at $N=60$ reflects the end of inflationary phase.}
    \label{Fig2:DynamicQuantitiesVsNe}
\end{figure}

\indent We consider a spatially flat FLRW metric with a Hubble rate $H=\dot a/a$, a homogeneous time-dependent scalaron field $\varphi(t)$ with potential energy $V(\varphi)$, and a radiation bath with temperature-dependent density $\rho_r(T)$. In WI, the scalaron interacts with the thermal bath through a dissipation coefficient $\Upsilon(\varphi,T)$. Motivated by warm little inflation realizations \cite{bastero-gil_warm_2016}, we focus only on the linearly temperature dependent dissipation coefficient $\Upsilon(T) = C_T T$ where $C_T$ encodes the microphysical strength of the dissipative interactions. Following refs.~\cite{freese_dark_2024, kamali_recent_2023, bastero-gil_warm_2016}, the general WI dynamic equations are,\\
\begin{align}
&H^2 = (\rho_\varphi+\rho_r)/(3M^2_{\rm Pl}), \label{F1}\\
&\dot\rho_r + 4H\rho_r = \Upsilon\,\dot\varphi^{\,2}, \label{rhorad}\\
&\ddot\varphi + 3H\dot\varphi + \partial V(\varphi)/ \partial \varphi = -\Upsilon\dot\varphi, \label{phiEOM}   
\end{align}
where $\rho_\varphi = V(\varphi)+\dot\varphi^{\,2}/2$, $\rho_r(T) = (\pi^2/30)g_\star T^4$ with $g_{\star}=106.75$ representing the number of relativistic species of the Standard Model (SM). Furthermore, we define two indispensable quantities, namely, the dissipation strength, 
\begin{align}
    Q = \frac{\Upsilon}{3H}, \label{Qparam}
\end{align}
that parametrizes the efficiency at which the scalaron converts into radiation and the HS potential energy (scalaron's potential energy) in a minimal coupling setting \cite{samart_warm_2022},
\begin{align}
    V(\varphi) = \frac{3}{4}\,M^2 M_{\rm Pl}^2\left(1-e^{-\sqrt{\frac{2}{3}}\frac{\varphi}{M_{\mathrm{Pl}}}}\right)^2
    \label{HSPotential}
\end{align}
where $M$ is the mass of scalaron in $M_{\mathrm{Pl}}$ units. Throughout this article, we consider only two scenarios with initial dissipation strengths $Q_0=\{10^{-2},1\}$. \\
\indent In WI dynamics, the primordial power spectrum is $\Delta^2_\mathcal{R}\propto G(Q)$ with $G(Q)$ being a growth function \cite{hall_scalar_2004, montefalcone_observational_2023, montefalcone_warmspy_2024, berera_identifying_2018, bastero-gil_shear_2011, bastero-gil_warm_2016, ghorui_warm_2025}. We fitted a $G(Q)=1 + 0.335 (Q^{1.364}) + 0.0185(Q^{2.315})$ in agreement with previous numerical calculations \cite{berera_identifying_2018, bastero-gil_shear_2011, bastero-gil_warm_2016, ghorui_warm_2025}.
Then the following inflationary observables were calculated, the primordial curvature perturbations  ${\displaystyle \ln(10^{10}A_{s})}=3.048$, the spectral index $n_s=0.9679$, and the tensor-to-scalar ratio $r=1.541\times10^{-4}$ in agreement with CMB values within $68\%$ CL \cite{planck_collaboration_planck_2020}. Finally, the scalaron mass $M=2.13\times10^{-6}  M_{\mathrm{Pl}}$ is fixed such that it fits the primordial curvature perturbations ${\displaystyle \ln(10^{10}A_{s})}$.\\
\indent The DM production depends on the evolution of $\{H(N_e),T(N_e), \rho_\varphi/\rho_r(N_e)\}$ parameters. In the slow--roll approximation, the parameter $\varepsilon_H = -\dot H/H^2$ is used to diagnose the WI dynamics. At $N_e = 60$, we obtain $\varepsilon_H = 1$, thus signaling the termination of inflation. In Fig.~\ref{Fig2:DynamicQuantitiesVsNe}, we show the evolution of the dynamic quantities in the $N_e$ basis when varying the dissipation strength. For $N_e\lesssim55$--$58$ the WI is in a quasi--steady state where the expansion is vacuum--driven ($\rho_\phi/\rho_r\gg1$), but dissipation continuously leaks the scalaron's kinetic energy to the thermal bath so that a $\rho_r\neq0$ is maintained. The collapse of $\rho_\varphi/\rho_r$ at $N_e\approx60$ reflects the onset of radiation era.
Physically, this means that the inflation proceeds without supercooling i.e., the bath temperature remains large enough ($T\gg H$) for thermal effects and dissipative friction to be relevant. In  Fig.~\ref{Fig2:DynamicQuantitiesVsNe}, the ratio $T/H\gg1, \textrm{for all}\, N_e$ signifies that the system is warm enough for WI to occur. In fact, $T/H\sim10-100$ in agreement with Ref.~\cite{kamali_recent_2023}. Having the values of $T/H$, we can give an estimate of $C_T=3HQ_0/T \leq 0.03$ for both weak and strong dissipation regimes within the inflationary phase. The values of $C_T$ are in agreement with Ref.~\cite{bastero-gil_warm_2016}.
Markedly, when $N_e\sim60$, the HS potential ceases to support accelerated expansion, the scalaron accelerates as it rolls off the plateau toward the minimum and its dissipation mechanism transfers sufficient energy such that radiation can occur. This is demonstrated at the bottom panel of Fig.~\ref{Fig2:DynamicQuantitiesVsNe}, when a sharp growth of $\varepsilon_H(N_{\textrm{end}}=60)=1$ happens, signaling further the termination of inflation. Right after the $\varepsilon_H=1$ moment, the radiation dominated (RD) phase smoothly starts, at $N_{\textrm{RD}} = 60.2\;\textrm{and}\;60.76$ for weak and strong dissipation regimes, respectively. During the RD era, $\varepsilon_H(N_{\textrm{RD}})\approx 2$ within our numerical tolerance. To understand why $\varepsilon_H(N_{\textrm{RD}})=2$ and stays flat after the RD transition we use Eqs.~\ref{F1}~--~\ref{rhorad}. From $\varepsilon_H=(3\dot \varphi^2+4\rho_r)/2(\rho_\varphi+\rho_r)$, we deduce that $\varepsilon_H(N_{\textrm{RD}})\approx 2$ when $\rho_r\gg \rho_\varphi$ implying a negligible kinetic term ($\dot\varphi\to0$). Concerning the late-time scaling of standard RD cosmology: $T\propto a^{-1}$ and $H\propto a^{-2}$ imply $Q\propto\Upsilon/H\propto a$, explaining the growth of $Q$ [top panel, Fig.~\ref{Fig2:DynamicQuantitiesVsNe}] after the exit ($N_e>60$). \\
\indent The dynamic quantities within the HS potential follow a similar behaviour to that in Ref.~\cite{freese_dark_2024}, despite the distinct form of the potential. However, our case exhibits one notable difference that is due to the flatness of the HS potential in the inflationary regime. Specifically, the transition between strong and weak dissipation regimes remains less sharply defined compared to steeper potentials like the quartic case \cite{freese_dark_2024}. This smoother transition arises because the plateau of HS potential leads to gradual change in the scalaron velocity ($\dot \varphi$), and consequently, in the dissipative dynamics that govern the thermal bath. While both dissipation regimes are still viable, the smooth transition means that intermediate dissipative behavior plays a significant role throughout the inflationary evolution. The dynamic quantities $\{H(N_e),T(N_e), \rho_\varphi/\rho_r(N_e)\}$ are the essential inputs for the DM yield as discussed in the next section.
\section{Dark Matter Production}\label{secDMyield}
In the UV freeze--in mechanism, it is assumed that DM is a stable relativistic species ($\chi$) with
no direct coupling to the scalaron ($\varphi$). Instead, $\chi$ interacts with the thermal bath through a non-renormalizable operator suppressed by a cutoff length scale $\Lambda$ \cite{freese_dark_2024, elahi_ultraviolet_2015, bernal_dawn_2017}. The nature of the DM particle and its stabilizing symmetry is discussed in detail in Sec.~\ref{natureDM}.
The DM number density ($n_\chi$) satisfies the Boltzmann equation \cite{bernal_uv_2020},
\begin{equation}
    \dot n_\chi + 3H n_\chi = \frac{T^{2n+4}}{\Lambda^{2n}}
    \label{BoltzmannEq}
\end{equation}
where the left hand side provides the evolution of $n_{\chi}$ with the expansion of the Universe serving as the friction term, and the right hand side parametrizes the cutoff scale $\Lambda$ of the effective field theory with non-renormalizable interactions of mass dimension $D\ge5$ where $D=n+4$ with $n\in \mathbb{N} $ \cite{elahi_ultraviolet_2015}. The right-hand side of Eq.~(\ref{BoltzmannEq}), represents the thermally averaged production rate in the relativistic regime where exclusively 2-to-2 scattering of SM particles occurs. This is dubbed the direct decay mechanism \cite{bernal_uv_2020, elahi_ultraviolet_2015}. Moreover, in this mechanism the DM mass has a lower bound of $m_\chi \sim 3$ keV. The condition $T/\Lambda \ll 1$, holds for all $\{N_{e},n\}$, and interactions of mass dimension $D>10$ are disfavored \cite{elahi_ultraviolet_2015,bernal_uv_2020}. These three conditions are successfully met within our framework.\\
\indent Using the DM yield $Y_\chi= n_\chi/s$ function (where the entropy $s=(2\pi^2/45)g_{\star,s}T^3$ with $g_{\star,s}$ being the effective number of relativistic species contributing to the SM entropy), and the e-fold definition $dN_e=Hdt$ we recast Eq.~\eqref{BoltzmannEq} into the known form \cite{freese_dark_2024},
\begin{align}
    Y_\chi(N_e) &= \frac{45}{2\pi^2 g_{\star,S}}\,\frac{e^{-3N_e}}{T^3(N_e)}
    \int_{N_{e,0}}^{N_e} I_\chi(N'_e) dN_e', \label{DMYieldInt}\\[4pt]
    I_\chi(N_e) &= \frac{e^{3N_e}}{H(N_e)}\frac{T^{2n+4}(N_e)}{\Lambda^{2n}},
    \label{IchiDef}
\end{align}
where $N_{e,0}$ denotes the initial value in which $Y_\chi(N_{e,0})=0$ is set. In order to produce the DM relic abundance $\Omega_\textrm{CDM}h^2=0.12$ \cite{planck_collaboration_planck_2020}, the DM mass ($m_\chi$) is fixed by the yield as follow:  $m_\chi Y_0=\Omega_\textrm{CDM} h^2 (\rho_c/s_0h^2)$ where $\rho_c\simeq1.1\times10^{-5} h^2$ GeV/cm$^3$ is the critical density and $s_0\simeq 2.91\times10^{3}$ cm$^{-3}$ is the present entropy density \cite{bernal_uv_2020}. \\
\indent To quantify the enhancement from WI dynamics, one can compare the final yield $Y_{\chi,\infty}$ obtained from Eq.~\eqref{DMYieldInt} to the conventional RD UV freeze--in yield ($Y_{\chi,\infty}^{\mathrm{RD}}$), defined as in \cite{freese_dark_2024}, 
\begin{align}
    Y_{\chi,\infty}^{\mathrm{RD}}(T_{\mathrm{rh}})\simeq
\frac{1}{\sqrt{2}}
\left(\frac{45}{\pi^{2}g_{\star}}\right)^{3/2}
\frac{1}{2n-1}\,
\frac{M_{\mathrm{Pl}}\,T_{\mathrm{rh}}^{2n-1}}{\Lambda^{2n}},
\end{align}
where it is evaluated at the same reheat temperature $T_{\rm rh}(N_{\textrm{RD}})\approx T(\varepsilon_H=2)$. 
Last but not least, we can rewrite Eq.~(\ref{IchiDef}) as the rate of change of the comoving number density $N_\chi= e^{3N_e}n_\chi$, i.e.\ $I_\chi = dN_\chi/dN_e$. These quantities are useful for a constructive discussion of DM yield. We describe the DM yield for two DM masses $m_\chi=1$ MeV and $m_\chi=100$ GeV. We want to emphasize that our study can be extended well beyond the actual studied DM mass range, DM mass range from $\mathcal{O}(\rm keV)$ to $\mathcal{O}(\textrm{PeV})$ \cite{bernal_uv_2020}, however the reason to choose the actual DM mass range was to compare this work with Ref.~\cite{freese_dark_2024}. Systematic calculations of the DM yield over the mass range $m_\chi\in[1\,\textrm{keV}, 1\,\textrm{TeV}]$ are provided in the Supplementary Material.
In the following section, the RD era is also benchmarked into the calculations of the DM yield. 
\subsection{Dark Matter with mass $m_\chi=1$ MeV}
\begin{figure}[!t]
    \centering
    \includegraphics[width=1\linewidth]{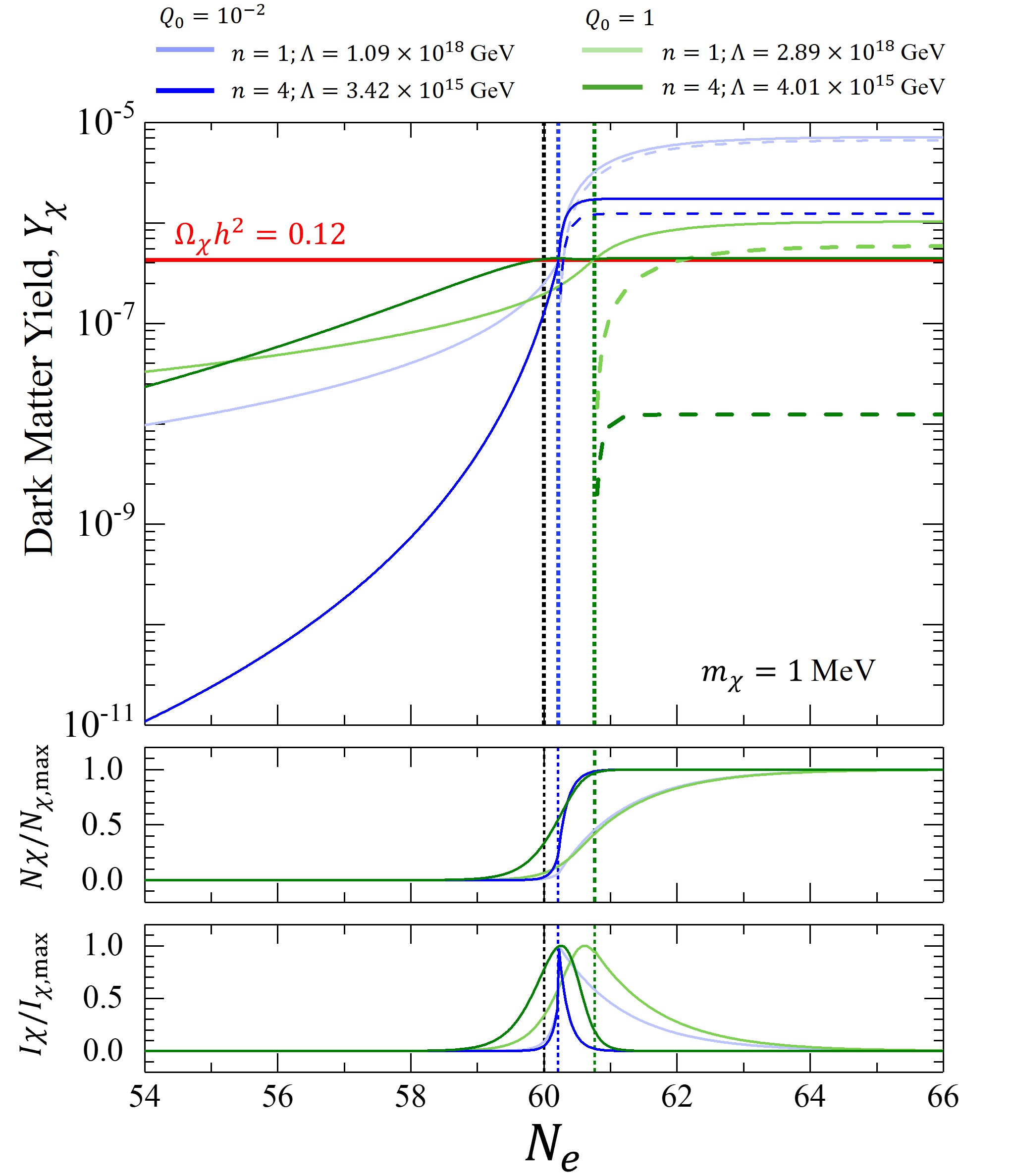}
    \caption{Top Panel: DM yield ($Y_\chi$) as a function of the e-folding for a DM mass $m_\chi=1$ MeV. In solid blue curves, the WI mechanism is described where in light blue (blue) color the weak regime $Q_0=10^{-2}$ for $n=1$ ($n=4$) is shown, respectively. In solid green curves, the WI mechanism is described where in light green (green) color the strong regime $Q_0=1$ for $n=1$ ($n=4$) is shown, respectively. 
    In dashed lines the RD onset is shown. The red horizontal line is a guide to the eye for the present-day observed DM relic abundance $\Omega_\textrm{CDM}h^2=0.12$ \cite{planck_collaboration_planck_2020}.
    Bottom Panel: Normalized comoving DM number density $N_\chi$ and its rate $I_\chi$ are plotted. In both panels, the vertical dotted lines signify the end of inflation.}
    \label{Fig3:DMYield1MeV}
\end{figure}

The DM yield via UV freeze--in in a direct decay mechanism within the HS potential is studied. In the top panel of Fig.~\ref{Fig3:DMYield1MeV}, we observe that the interplay between the strong dissipation regime $Q_0=1$ and $n=4$ $(D=8)$ produces the DM relic abundance ($\Omega_\textrm{CDM}h^2=0.12$) close to the termination of the inflation $N_e\approx 59.5-60$ and then smoothly continues for $N_e>60$. Once the transition at $N_e=60$ occurs, the RD takes over. Due to the DM mass choice we made, the RD era given by $\{Q_0=1,n=1\}$ ($D=5$) overestimates the DM relic abundance. Recall that, in UV freeze--in the DM mass $m_\chi$ does not enter the yield $Y_{0}$; it appears only in the DM relic abundance $\Omega_{\mathrm{CDM}} h^2=m_\chi Y_{0} s_0 / \rho_c$. A factor of $\sim2$ lighter mass (e.g. $\sim 0.5 \,\mathrm{MeV}$ ) reproduces exactly the relic abundance without altering the dynamics. We retain 1 MeV for comparison with Ref.~\cite{freese_dark_2024}.
\\
\indent This extension to the RD era is made visible from $N_\chi$ [see bottom panel of Fig.~\ref{Fig3:DMYield1MeV}]. The light green curve $\{Q_0=1,n=1\}$ ($D=5$) starts to rise close to the termination of inflation and evolves through the RD era. In contrast, the green curve $\{Q_0=1,n=4\}$ ($D=8$) has an earlier rise and evolves right after $N_e=60$, while the weak dissipation regime (blue curves in Fig.~\ref{Fig3:DMYield1MeV}) shows a higher overestimation of the DM relic abundance. The rate of change $I_\chi$ given in the bottom panel distinguishes the peaks where the transition from inflation to RD era is occurring. In the considered cases of Fig.~\ref{Fig3:DMYield1MeV}, the legend shows how fixing $\Omega_\textrm{CDM}h^2=0.12$ translates into microphysical scales: for $n=1$ one needs $\Lambda\sim10^{18}\,\mathrm{GeV}$ (Planckian scale), varying only by a factor of a few between weak ($Q_0=10^{-2}$) and strong ($Q_0=1$) dissipation regimes, whereas for $n=4$ the required scale is lower, $\Lambda=4.01\times10^{15}\,\mathrm{GeV}$, showing the enhanced efficiency of strongly-temperature dependent DM production near the RD era. In summary, UV freeze--in in warm HS inflation is dominated by a short e-folding window ($\Delta N_e$) where inflation$\to$radiation transition occurs, with $n$ controlling the DM yield and $Q_0$ the shifting of the DM yield. The $\Lambda$ needed to reproduce the present--day DM relic abundance $\Omega_\textrm{CDM}h^2=0.12$ is $\Lambda=4.01\times10^{15}\,\mathrm{GeV}$ with $\{Q_0=1,n=4\}$. As we show in the next section, $Q_0=1$ also favors the DM mass $m_\chi=100$ GeV.
\begin{figure}[!t]
    \centering
    \includegraphics[width=1\linewidth]{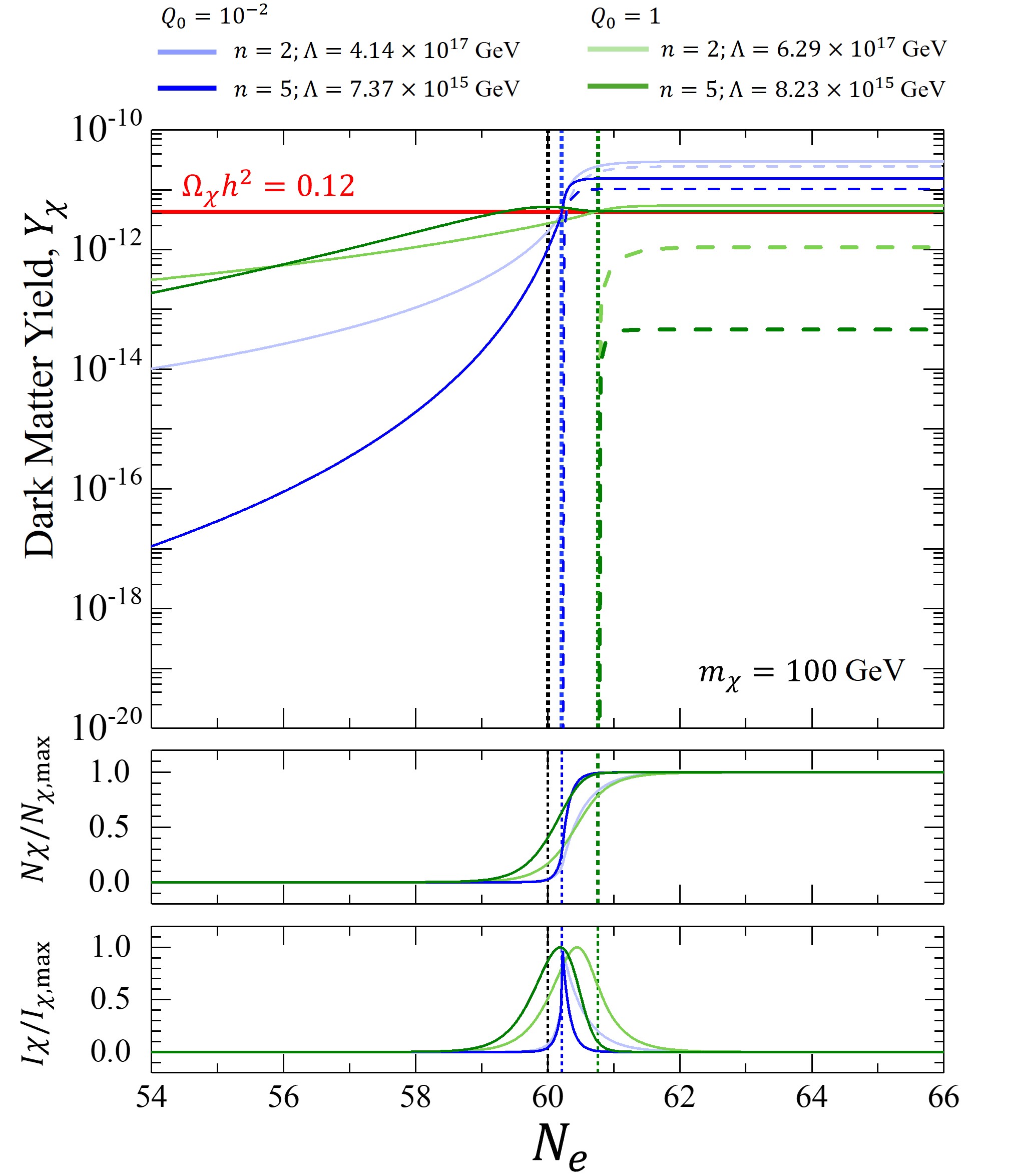}
    \caption{Top Panel: DM yield ($Y_\chi$) as a function of the e-folding for a DM mass $m_\chi=100$ GeV. In solid blue curves, the WI mechanism is described where in light blue (blue) color the weak regime $Q_0=10^{-2}$ for $n=2$  ($n=5$) is shown, respectively. In solid green curves, the WI mechanism is described where in light green (green) color the strong regime $Q_0=1$ for $n=2$    ($n=5$) is shown, respectively. 
    In dashed lines the RD onset is shown. The red horizontal line is a guide to the eye for the present-day observed DM relic abundance $\Omega_\textrm{CDM}h^2=0.12$ \cite{planck_collaboration_planck_2020}.
    Bottom Panel: Normalized comoving DM number density $N_\chi$ and its rate $I_\chi$ are plotted. In both panels, the vertical dotted lines signify the end of inflation.}
    \label{Fig4:DMYield100GeV}
\end{figure}\\

\subsection{Dark Matter with mass $m_\chi=100$ GeV}
The DM yield for DM mass $m_\chi=100$ GeV is studied and results are given in Fig.~\ref{Fig4:DMYield100GeV} for strong and weak dissipation regime. In particular, for the strong dissipation case ($Q_0=1$), the WI mechanism provides sufficient energy transfer into the thermal bath to reproduce the observed DM relic abundance, with the DM yield rising rapidly and approaching the target horizontal line $\Omega_\textrm{CDM}h^2=0.12$ near the transition $N_e\simeq58$--$61$ [top panel of Fig.~\ref{Fig4:DMYield100GeV}]. After the transition, the DM yield continues to evolve smoothly entering into the RD era for $N_e >61$, although its growth becomes increasingly suppressed as the thermal bath cools and the production rate drops. This feature is important in assessing simplified approximations: the RD benchmark curves shown as dashed lines in the top panel of Fig.~\ref{Fig4:DMYield100GeV} do not fully capture the DM production across the transition. Concretely, they underestimate the final DM yield in the strong dissipation regime (green and light green solid curves), whereas a non-negligible fraction of the DM relic abundance is accumulated just after the end of inflation, while they overestimate the DM abundance in the weak dissipation regime (blue and light blue curves), where production is sharply localized near the transition and shuts off rapidly. \\
\indent The normalized comoving DM number density ($N_\chi$) plotted at the bottom panel of Fig.~\ref{Fig4:DMYield100GeV}, exhibits a rapid increase from $0$ to $1$ around the inflation--radiation transition. This shows that essentially all of the final DM yield is accumulated within a narrow $\Delta N_e\sim 2-4$ e-folding window. Consistently, the normalized $I_\chi$ peaks sharply near $N_e\simeq 58$--$61$, identifying this epoch as the dominant DM production window; larger mass dimensions $D$ lead to narrower and taller peaks, while the strong dissipation case ($Q_0=1$) slightly broadens and shifts the DM production window compared to the weak dissipation ($Q_0=10^{-2}$).

\indent Overall, we demonstrate that the WI UV freeze--in within HS comprises a viable mechanism to produce the DM relic abundance. The observed DM relic abundance can be reproduced by choosing $\Lambda$, where heavier DM mass require lower $\Lambda$ (e.g. $n=5$ giving $\Lambda\sim10^{15}\,\mathrm{GeV}$ versus $n=2$ giving $\Lambda\sim10^{17}\,\mathrm{GeV}$). The strong dissipation regime ($Q_0=1$) robustly achieves $\Omega_\textrm{CDM}h^2=0.12$ for both $n=\{2,5\}$, while also highlighting that the simplified RD benchmarks may misestimate the final DM yield across the transition, underscoring the importance of evolving the full WI dynamics through the termination of inflation. The DM mass is one out of three important ingredients to describe the nature of the particle. The two other ingredients, charge and spin, are discussed in the next section.

\section{Dark Matter Candidates: Symmetries and Operator Classification}
\label{natureDM}
The UV freeze--in mechanism within warm inflation dynamics is, by construction, independent of the DM spin \cite{elahi_ultraviolet_2015, bernal_uv_2020}. In other words, the DM yield has the same scaling law $Y\sim M_{\textrm{Pl}}T^{2n-1}/\Lambda^{2n}$ (up to a $\mathcal{O}(1)$ correction factor) when either a scalar, fermion or a vector boson state is assumed \cite{elahi_ultraviolet_2015}. We observe that the Boltzmann collision term in Eq.~\ref{BoltzmannEq} is dependent only on the temperature and the mass dimension of the effective cutoff scale term $\Lambda$ that connects the dark and visible sectors. Additionally, as refs.~\cite{bernal_uv_2020, bernal_dawn_2017, elahi_ultraviolet_2015} for the UV freeze--in mechanism to be consistent, the cutoff scale should be the highest scale in the calculation i. e., $m_{\chi}\ll T_{\textrm{rh}}\ll \Lambda$. \\
\indent The analysis performed in Sec. 2 and 3 constitutes a general framework for arbitrary spin. However, in this section discuss which shape of the non-renormalizable interactions can give the dominance of $D=8$ operator for 1 MeV of DM mass [Fig.~\ref{Fig3:DMYield1MeV}] and the dominance of $D=9$ operator for 100 GeV of DM mass [Fig.~\ref{Fig4:DMYield100GeV}]. Moreover, we discuss the particle nature of DM, its stabilizing symmetry and if any extra symmetries is required for explaining the dominance of the aforementioned high dimensional operators. \\
\indent The minimal choice for the DM to be a stable particle is to impose the $\mathbb{Z}_2$ symmetry \cite{hall_freeze-production_2010, elahi_ultraviolet_2015, Barman_2020, barman2021, song2023completeeftoperatorbases}. The $\mathbb{Z}_2$ symmetry guarantees the stability of the DM particle in the absence of a vacuum expectation value (VEV). Additionally, the $\mathbb{Z}_2$ symmetry forbids all operators linear in the DM field. Subsequently, we denote $\phi$ as the real scalar DM singlet, $\rho$ as the DM fermion, and $X_\mu$ as the DM vector. Two model-building facts then organize the
operator content. First, field bilinear with even mass dimension (both bilinear scalars and vector bosons having dimension $2$), whereas the fermion bilinear has odd mass dimension $3$. This single parity difference dictates which total mass dimension $D$ each spin can reach with baryon-lepton $B-L$ conserving SM operators \cite{Kobach_2016, Helset_2020}. Second, there is no $D=1$ SM-field operator to accompany the scalar bilinear $\phi^2$: the lowest SM gauge-invariant scalar is $H^\dagger H$ (dimension $2$) where $H$ is the SM Higgs doublet. Consequently, the bosonic $\Lambda$ tower starts at $D=4$ ($\phi^2 H^\dagger H$) and the fermionic tower starts at $D=5$ $(\rho^2_c H^\dagger H)$. To further elucidate the discussion, we conside the effective Lagrangian, 
\begin{align}
    \mathcal{L}_{\mathrm{eff}}=\mathcal{L}_{\mathrm{SM}}+\mathcal{L}_{\textrm{DM}}+ \frac{1}{\Lambda^{D-4}}\left(\sum_i \mathcal{O}_i^{(D)}+\text {h.c.}\right)
\end{align}
where $\mathcal{L}_{\mathrm{SM}}$ is the renormalizable SM Langrangian with gauge structure ($\mathrm{SU}(3)_c \times \mathrm{SU}(2)_L \times \mathrm{U}(1)_Y$), $\mathcal{L}_{\mathrm{DM}}$ is the DM Lagrangian to be assumed and the last term is the non-renormalizable Lagrangian with each operator $\mathcal{O}$ is suppressed by $1/\Lambda^{(D-4)}$. The DM Lagrangian reads as, 
\begin{align}
\mathcal{L}_{\mathrm{DM}}=\left\{\begin{array}{l}
\frac{1}{2} \partial_\mu \phi \partial^\mu \phi-\frac{1}{2} m_\phi^2 \phi^2 - \frac{\lambda}{4!}\phi^{4}\\
i \bar{\rho} \slashed{\partial} \rho-m_\rho \bar{\rho} \rho \\
-\frac{1}{4} X_{\mu \nu} X^{\mu \nu}+\frac{1}{2} m_X^2 X_\mu^2
\end{array}\right.
\label{LagrangianDM}
\end{align}
where $\{m_\phi,\,m_\rho\,,m_X\}$ are the DM masses, and $X_{\mu \nu}$ is the vector DM field strength tensor. Scalar and fermionic DM fields preserve the SM gauge structure while a vector DM field requires the introduction of an extra symmetry e. g. changing the SM gauge structure to $\mathrm{SU}(3)_c \times \mathrm{SU}(2)_L \times \mathrm{U}(1)_Y \times \mathrm{U}(1)'$ \cite{elahi_ultraviolet_2015}. In this work, we restrict our discussion to fields that do not require an extension of the SM gauge structure thus we avoid the vector DM boson. Moreover, we preserve the common definitions of $B_{\mu\nu}$ as the hypercharge field strength, $W^j_{\mu\nu}$ as the $SU(2)_{L}$ gauge boson, $G^a_{\mu\nu}$ as the gluon field, $\psi$ as a generic SM multiplet. Let us now consider the scalar and fermionic DM states. \\
\indent Scalar Dark Matter ($\phi$): Consider the case where $\phi$ is 
a real scalar singlet field stabilized by a $\mathbb{Z}_2$. We focus exclusively 
on the non-renormalizable operators, leaving the renormalizable SM and scalar DM 
Lagrangian [Eq.~\ref{LagrangianDM}] implicit. An appealing case is the dominance 
of the $n=5$ $(D=9)$ operators, whose contribution rise before the end of inflation and 
reproduce the correct relic abundance in the RD era with cutoff scale 
$\Lambda=8.23\times10^{15}$~GeV [Fig.~\ref{Fig4:DMYield100GeV}]. For the $D=9$ 
operators to dominate, the $D\leq8$ operators must be suppressed or forbidden by 
an additional symmetry while still preserving the SM gauge structure. Assigning $\phi$ a 
lepton number \cite{gardner_2020} forbids the $B-L$ conserving even-dimensions: because the scalar bilinear $\phi^2$ has even mass dimension, the surviving $\Delta L=2$ 
operators built on the Weinberg structure $(LH)(LH)$ appear at the odd 
dimensions $D=\{7,9\}$. We list a few representative $D=9$ operators: $(D_\mu)^2\phi^2(LH)(LH)$, $\phi^4(LH)(LH)$, $\phi^2(LH)(LH)H^\dagger H$, and $\phi^2(LL)H^2 F_L$, where $F=\{B,W,G\}$ are the SM gauge field strengths, $L$ the lepton doublets, and the tensor bilinear is contracted with $F_{L,\mu\nu}$. Note that the leading $D=7$ Weinberg operator $\phi^2(LH)^2$ shares the quantum numbers of the $D=9$ operators \cite{Li_2021}, so $D=9$ operator dominance is mainly controlled by the UV completion rather than by an exact symmetry. Here, there is an attractive coincidence of scales: the $D=5$ Weinberg operator evaluated at this cutoff would generate: 
$m_\nu\simeq v^2/\Lambda\sim(246\,\textrm{GeV})^2/(8.23\times10^{15}\,\textrm{GeV})\sim 7\times10^{-3}$~eV, the observed neutrino-mass scale (where $v$ is the Higgs VEV). The $D=9$ operator itself generates a parametrically smaller neutrino mass contribution, so $\phi$ ties the dark sector to the neutrino sector through a shared $\Delta L=2$ structure. Scalars carrying lepton number can participate in $B-L$ violating processes such as neutrinoless double-$\beta$ decay \cite{Kobach_2016, Helset_2020, gardner_2020, Babu_2001}. A second, more direct realization of this connection to neutrino mass generation is provided by the $D=8$ operator in the fermionic DM case discussed below.\\
\indent Fermionic Dark Matter $(\rho)$: Consider the case where the DM 
particle is a fermion gauge singlet $(\rho)$ with $n_\rho$ flavors. Imposing a 
$\mathbb{Z}_2$ symmetry makes the DM particle stable and forbids $D=4$ operators 
\cite{song2023completeeftoperatorbases}. As shown in Fig.~\ref{Fig3:DMYield1MeV}, the relic abundance is reproduced by operators of mass dimension $D=8$ at the inflation-to-radiation transition, with a cutoff scale $\Lambda=4.01\times10^{15}$~GeV. For the $D=8$ operators to dominate, the 
$D\leq7$ operators must be suppressed or forbidden by an additional symmetry 
while preserving the SM gauge structure. Assigning a lepton number to $\rho$ that 
forbids the $B-L$ conserving portals achieves this: because the fermion 
bilinear $\rho_c^2$ has odd mass dimension, the unique $\Delta L=2$ operator $H^2 L^2 \rho_c^2$, sits at the even dimension $D=8$ in contrast to the scalar case, 
where the $\Delta L=2$ operators appeared at odd dimensions $D=7,9$. Representative $D=8$ operators include the $B-L$ conserving $(D_\mu) H^2 H^{\dagger 2}\rho_c\rho_c^\dagger$ and $H H^\dagger L L^\dagger \rho_c\rho_c^\dagger$, together with the $B-L$ violating $H^2 L^2\rho_c^2$ operator \cite{song2023completeeftoperatorbases, barman2021}. Upon an electroweak symmetry breaking, the operator  $H^2 L^2\rho_c^2$ might contribute to the neutrino mass; for comparison, the $D=5$ Weinberg operator at the same scale would yield $m_\nu\simeq v^2/\Lambda\sim(246\,\textrm{GeV})^2/(4.01\times10^{15}\,\textrm{GeV}) \sim 1.5\times10^{-2}$~eV, consistent with the observed neutrino mass scale. The $D=8$ operator generates a parametrically smaller contribution, suppressed by additional powers of $v/\Lambda$, so that the fermionic DM candidate ties the dark sector to the origin of neutrino masses while leaving the leading Weinberg contribution intact. 

\section{Discussion and Summary}
\label{discSumm}
In this work, we investigated scalaron--driven DM production during warm inflation via UV freeze--in mechanism. The scalaron potential energy is relatively insensitive to the two initial dissipation strengths [Fig.~\ref{Fig2:DynamicQuantitiesVsNe}]. This feature leads to similar thermal histories for both weak and strong dissipation regimes until the termination of inflation. Consequently, DM production via UV freeze--in is strongly localized around the inflation--to--radiation transition, where the temperature remains high enough.
For DM with mass $m_\chi=1$ MeV [Fig.~\ref{Fig3:DMYield1MeV}], we demonstrated that the present-day DM relic abundance $\Omega_\textrm{CDM}h^2=0.12$ can be achieved for non-renormalizable interactions with $n=1$ (RD, weak dissipation regime) and $n=4$ (inflation--to--radiation transition, strong dissipation regime) [Fig.~\ref{Fig3:DMYield1MeV}]. The required cutoff scale $\Lambda$ spans from Planckian values $\mathcal{O}(10^{18})$ GeV for $n=1$ ($D=5$) down to $\mathcal{O}(10^{15})$ GeV for $n=5$ ($D=9$), reflecting the enhanced temperature dependence of higher dimensional operators. For DM with mass $m_\chi=100$ GeV [Fig.~\ref{Fig4:DMYield100GeV}], the UV freeze--in mechanism remains viable, particularly in the strong dissipation regime ($Q_0=1$), where sufficient energy is transferred to the thermal bath to sustain efficient production. In this case, essentially the entire DM abundance is generated within a narrow e--folding window ($\Delta N_e$) around the transition between inflation--radiation. In general, the strong dissipation regime robustly achieves $\Omega_\textrm{CDM}h^2=0.12$ for both $n=\{4,5\}$ $(D=\{8,9\})$, as shown by the sharp features in the comoving number density and production rate [Fig.~\ref{Fig4:DMYield100GeV}]. These results extend the work of Ref.~\cite{freese_dark_2024}. We want to emphasize that the warm HS inflation within the UV freeze--in mechanism depends only on the mass dimension $D$ of the non-renormalizable operator, where the nature of the DM particle enters only through $\mathcal{O}(1)$ prefactors that leave the temperature and the cutoff scaling unchanged. This framework provides a tool to identify, for arbitrary spin, which operator dimensions are accessible and how their contributions scale with the reheating temperature [Eq.~\ref{BoltzmannEq}]. An attractive coincidence was found for both scalad and fermionic DM particles: the cutoff scale of their leading non-renormalizable operator would generate the neutrino mass if evaluated at the $D=5$ Weinberg operator.
\\
\indent In conclusion, we demonstrated that scalaron-driven DM during Warm Inflation via UV freeze--in provides a consistent and efficient mechanism to produce the present-day DM relic abundance across a range of DM masses $m_\chi=\{1\,\textrm{MeV}, 100\,\textrm{GeV}\}$ through non-renormalizable interactions with $n=\{4,5\}$ (i.e., mass dimension $D=\{8,9\}$) at strong dissipation regimes $Q_0=1$. An extended systematic calculation of the DM yield for $m_\chi\in[1\,\textrm{keV}, 1\,\textrm{TeV}]$, illustrating the monotonic behavior of the UV freeze--in mechanism, is given in Figs. S1-S3 of the Supplementary Material.
The obtained results emphasize the significance of the inflation--to--radiation transition as a pivotal moment when DM production can occur and motivate further studies exploring different dissipation mechanisms \cite{berghaus_warm_2024} and interaction structures \cite{covi_axinos_2001, bae_effective_2011, volkas_naturalness_1988}.
\section*{Data availability Statement} 
No observational datasets were used. All numerical results were produced within the theoretical framework presented in this manuscript, and the data underlying the figures are available from the authors upon reasonable request.

\bibliography{sn-bibliography.bib}

\theoremstyle{thmstyleone}%
\theoremstyle{thmstyletwo}%
\newcolumntype{Y}{>{\raggedright\arraybackslash}X}
\theoremstyle{thmstylethree}%
\setcounter{figure}{0}
\renewcommand{\thefigure}{S\arabic{figure}}
\raggedbottom
\onecolumn


\section*{\centering Supplementary Note to Scalaron-driven Dark Matter during Warm Inflation via UV Freeze--in}
In this supplementary, we plot systematically the Dark Matter (DM) yield for $n=\{1,2,3,4,5\}$ (mass dimension $D=\{5,6,7,8,9\}$). The plots are shown for a DM mass range $m_{\chi}\in[1\,\rm{keV}, 1\,\rm{TeV}]$, larger than is discussed in the main text. Fig. S1 shows the calculations for mass dimension $D=\{5,6\}$, Fig. S2 for $D=\{7,8\}$, and Fig. S3 compares the lowest and highest operator mass dimensions discussed in the main text, $D=\{5,9\}$. For each mass dimensions, the corresponding cut-off scale is calculated. 
\section*{\centering DM Yield for $n=\{1,2\}$ (mass dimension $D=\{5,6\}$)}
\vspace{-5mm}
\begin{figure}[!htb]
    \includegraphics[width=\linewidth]{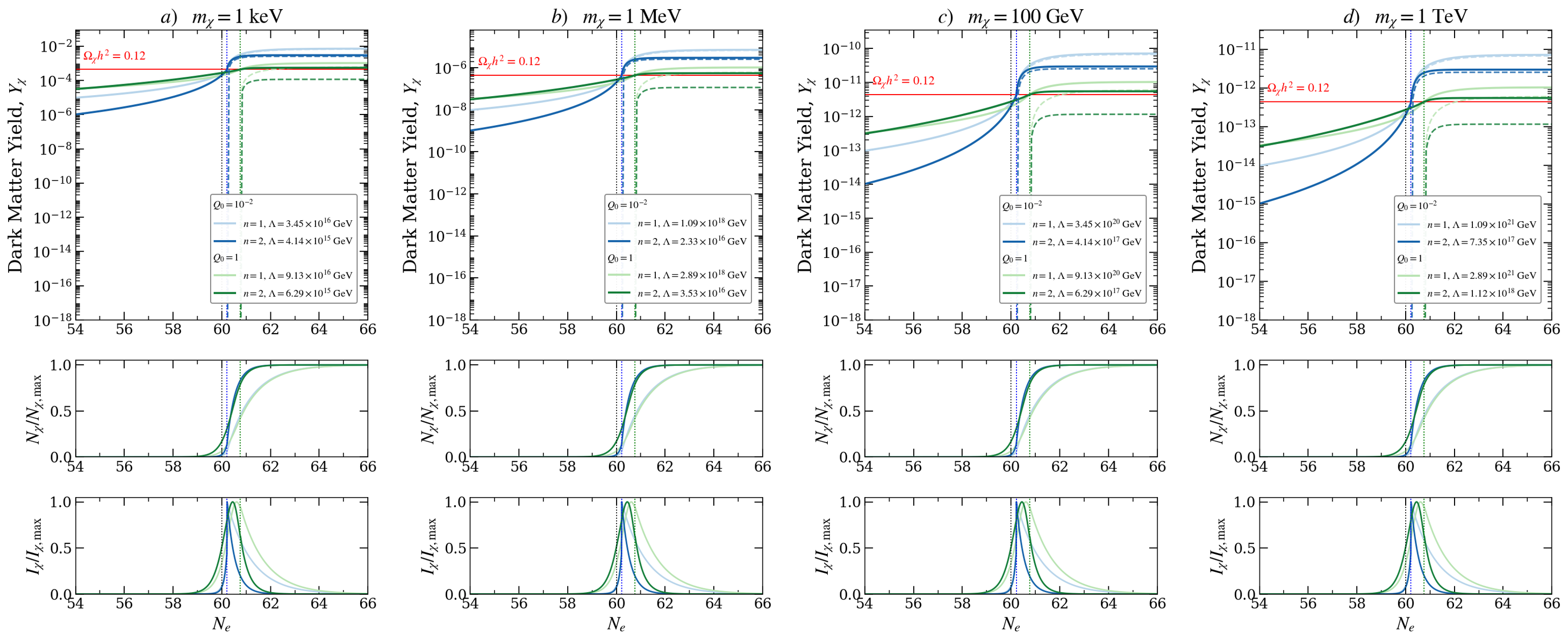}
    \caption{Top Panel: DM yield ($Y_\chi$) as a function of the e-folding ($N_e$) for a range of DM mass $m_{\chi}\in[1\,\rm{keV}, 1\,\rm{TeV}]$. In solid blue curves, the WI mechanism is described where in light blue (blue) color the weak regime $Q_0=10^{-2}$ for $n=1$ ($n=2$) is shown, respectively. In solid green curves, the WI mechanism is described where in light green (green) color the strong regime $Q_0=1$ for $n=1$ ($n=2$) is shown, respectively. 
    In dashed lines the RD onset is given. The red horizontal line is a guide to the eye for the present-day observed DM relic abundance $\Omega_\textrm{CDM}h^2=0.12$.
    Bottom Panel: Normalized comoving DM number density $N_\chi$ and its rate $I_\chi$ are plotted. In both panels, the vertical dotted lines signify the end of inflation.}
\end{figure}

\newpage
\section*{\centering DM Yield for $n=\{3,4\}$ (mass dimension $D=\{7,8\}$)}
\vspace{-5mm}
\begin{figure}[!htb]
    \includegraphics[width=\textwidth]{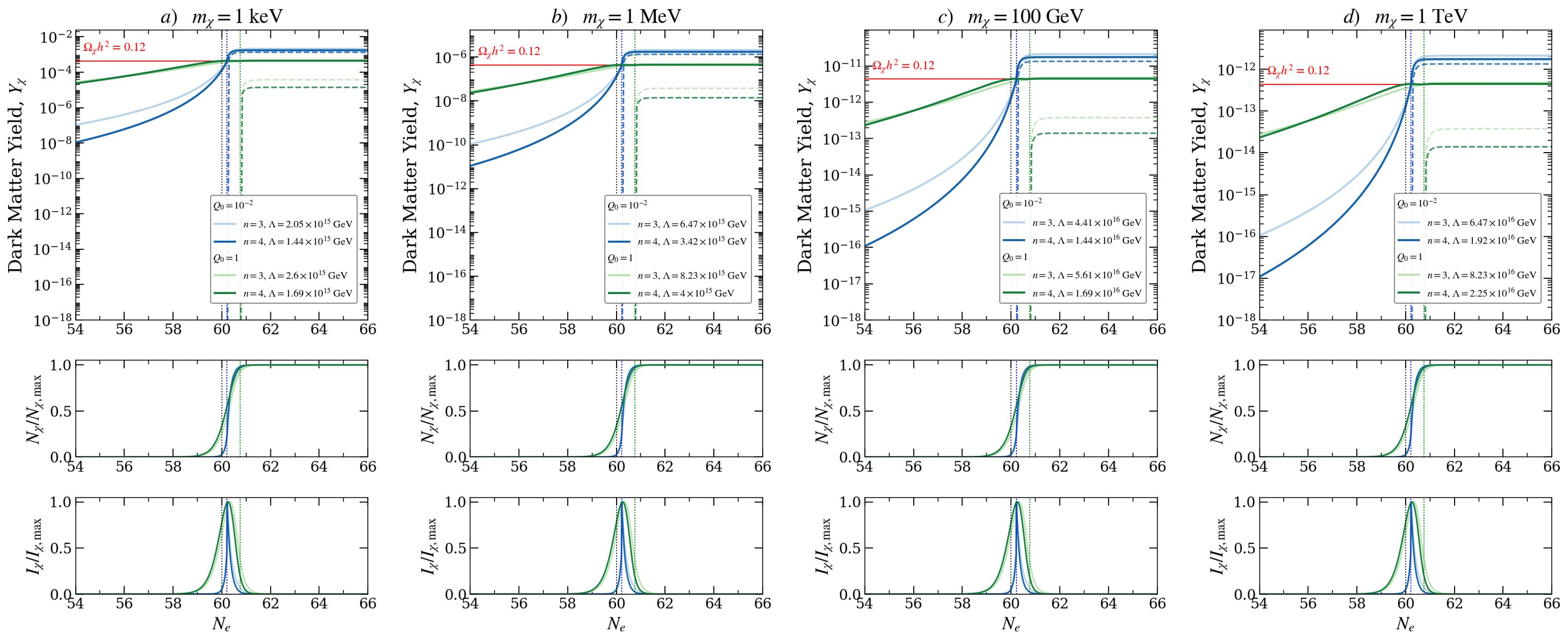}
    \caption{Top Panel: DM yield ($Y_\chi$) as a function of the e-folding ($N_e$) for a range of DM mass $m_{\chi}\in[1\,\rm{keV}, 1\,\rm{TeV}]$. In solid blue curves, the WI mechanism is described where in light blue (blue) color the weak regime $Q_0=10^{-2}$ for $n=3$ ($n=4$) is shown, respectively. In solid green curves, the WI mechanism is described where in light green (green) color the strong regime $Q_0=1$ for $n=3$ ($n=4$) is shown, respectively. 
    In dashed lines the RD onset is given. The red horizontal line is a guide to the eye for the present-day observed DM relic abundance $\Omega_\textrm{CDM}h^2=0.12$. 
    Bottom Panel: Normalized comoving DM number density $N_\chi$ and its rate $I_\chi$ are plotted. In both panels, the vertical dotted lines signify the end of inflation.}
\end{figure}

\section*{\centering DM Yield for $n=\{1,5\}$ (mass dimension $D=\{5,9\}$)}
\vspace{-5mm}
\begin{figure}[!htb]
    \includegraphics[width=\textwidth]{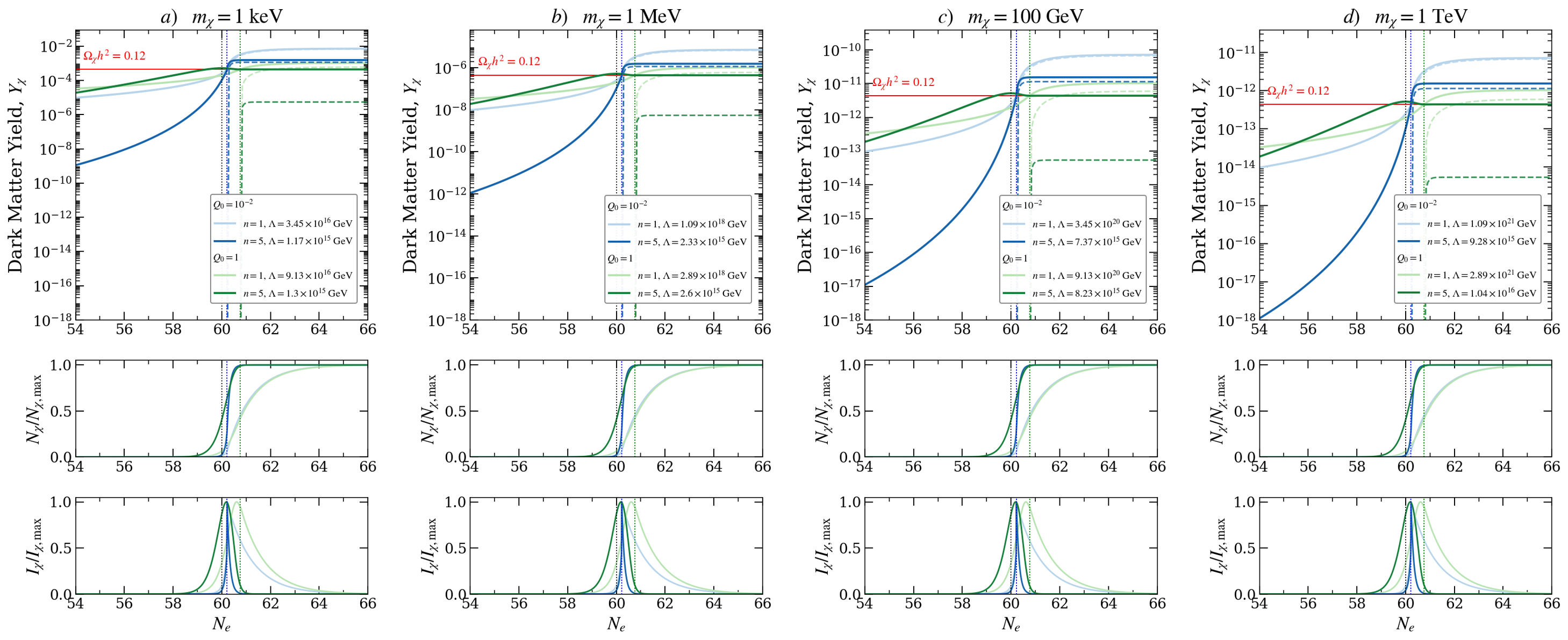}
    \caption{Top Panel: DM yield ($Y_\chi$) as a function of the e-folding ($N_e$) for a range of DM mass $m_{\chi}\in[1\,\rm{keV}, 1\,\rm{TeV}]$. In solid blue curves, the WI mechanism is described where in light blue (blue) color the weak regime $Q_0=10^{-2}$ for $n=1$ ($n=5$) is shown, respectively. In solid green curves, the WI mechanism is described where in light green (green) color the strong regime $Q_0=1$ for $n=1$ ($n=5$) is shown, respectively. 
    In dashed lines the RD onset is given. The red horizontal line is a guide to the eye for the present-day observed DM relic abundance $\Omega_\textrm{CDM}h^2=0.12$.
    Bottom Panel: Normalized comoving DM number density $N_\chi$ and its rate $I_\chi$ are plotted. In both panels, the vertical dotted lines signify the end of inflation.}
\end{figure}

\end{document}